\documentclass[letterpaper,twocolumn,9pt]{hotnets15-cr}

\usepackage{times}
\usepackage[TABBOTCAP]{subfigure}
\usepackage{tabularx}
\usepackage{graphicx}
\usepackage{color}
\usepackage{xspace}
\usepackage{thumbpdf}
\usepackage{listings}
\usepackage{verbatim}
\usepackage{booktabs}
\usepackage{colortbl}
\usepackage{url}
\usepackage{balance}

\usepackage{microtype}

\usepackage{cite}

\graphicspath{{./figs/}}

\conferenceinfo{HotNets '15}{November 16--17 2015, Philadelphia, PA USA}
\CopyrightYear{2015}
\crdata{978-1-4503-4047-2}
\doi{http://dx.doi.org/10.1145/2834050.2834102}

\newenvironment{senumerate}{%
   \begin{list}{\arabic{enumi}.}{%
    \setlength\labelwidth{1.5em}%
    \setlength\leftmargin{1.5em}%
    \setlength{\topsep}{4pt plus 2pt minus 2pt}%
    \setlength\itemsep{0.0cm}%
    \usecounter{enumi}}%
  }{\end{list}}

\newcommand{\eg}{\textit{e.g.}~}
\newcommand{\ie}{\textit{i.e.}~}

\newcommand{\one}{({\em i})\xspace}
\newcommand{\two}{({\em ii})\xspace}
\newcommand{\three}{({\em iii})\xspace}

\usepackage{pifont}
\newcommand{\cmark}{\ding{51}}%
\newcommand{\xmark}{\ding{56}}%
\newcommand{\pmark}{\ding{218}}%

\usepackage[absolute,showboxes]{textpos}

\TPMargin{5pt}

\newcommand{\copyrightstatement}{
    \begin{textblock}{0.8}(0.1,0.02)    
         \noindent
         \footnotesize
         If you cite this paper, please use the HotNets reference:
         M. W\"ahlisch, R. Schmidt, T.C. Schmidt, O. Maennel, S. Uhlig,
         G. Tyson, ``RiPKI: The Tragic Story of RPKI Deployment in the
         Web Ecosystem'', in \emph{Proc. of ACM HotNets}, ACM, 2015.
    \end{textblock}
}

\begin{document}
\clubpenalty=10000
\widowpenalty = 10000

\title{RiPKI: The Tragic Story of RPKI Deployment \\in the Web Ecosystem}

\numberofauthors{6}
\author{
\alignauthor Matthias W\"ahlisch\\
       \affaddr{Freie Universit\"at Berlin}\\
       \email{m.waehlisch@fu-berlin.de}
\and
\alignauthor Robert Schmidt\\
       \affaddr{Freie Universit\"at Berlin}\\
       \email{rs.schmidt@fu-berlin.de}
\and
\alignauthor Thomas C. Schmidt\\
       \affaddr{HAW Hamburg}\\
       \email{t.schmidt@haw-hamburg.de}
\and
\alignauthor Olaf Maennel\\
       \affaddr{Tallinn U. of Technology}\\
       \email{olaf.maennel@ttu.ee}
\and
\alignauthor Steve Uhlig\\
       \affaddr{Queen Mary Univ. London}\\
       \email{steve@eecs.qmul.ac.uk}
\and
\alignauthor Gareth Tyson\\
       \affaddr{Queen Mary Univ. London}\\
       \email{g.tyson@qmul.ac.uk}
}

\copyrightstatement
\maketitle

\begin{abstract}
Web content delivery is one of the most important services on the Internet.  Access to websites is typically secured via TLS. However, this security model does not account for prefix hijacking on the network layer, which may lead to traffic blackholing or transparent interception. Thus, to achieve comprehensive security and service availability, additional protective mechanisms are necessary such as the RPKI, a recently deployed \underline{R}esource \underline{P}ublic \underline{K}ey \underline{I}nfrastructure to prevent hijacking of traffic by networks.  This paper argues two positions. First, that modern web hosting practices make route protection challenging due to the propensity to spread servers across many different networks, often with unpredictable client redirection strategies; and, second, that we need a better understanding why protection mechanisms are \emph{not} deployed. To initiate this, we empirically explore the relationship between web hosting infrastructure and RPKI deployment. Perversely, we find that less popular websites are more likely to be secured than the prominent sites. Worryingly, we find many large-scale CDNs do not support RPKI, thus making their customers vulnerable. This leads us to explore business reasons why operators are hesitant to deploy RPKI, which may help to guide future research on improving Internet security.

\end{abstract}




\section{Introduction}
\label{sec:intro}

Website security is a long pursued and rather esoteric goal. Traditionally, it has been approached from an end-to-end perspective
(\eg TLS), largely because this is easily within the sphere of control of any web provider. However, as evidenced by many prominent
attacks, this is frequently insufficient. This is because various third party infrastructure dependencies exist that maybe vulnerable
to attack, \eg within BGP, DNS, certificate authorities, operating systems.

Take, for example, BGP, which manages inter-domain connectivity. If BGP is misconfigured, websites
can become severed from the network, regardless of the security measures taken at the application layer~\cite{youtube-outage,sb-arc-10,indonesia-hijack}.
Thus, true security requires multi-layer and multi-stakeholder cooperation. In this paper, we focus on the importance of routing layer security
for ensuring reliable website provision.

Routing security is a topic that is discussed widely among network operators, yet (anecdotally) receives far less attention
from web service providers. This is despite the availability of technologies that mitigate attacks. For example,
the Resource Public Key Infrastructure (RPKI) is a recently standardized
framework that enables BGP routers to perform prefix origin validation.
When activated, this can prevent incidents such as the Pakistan Telecom hijacking of
traffic destined to YouTube. It works by crytographically validating the right of an AS to advertise
a prefix. Although clearly attractive as a security measure, it is unclear to what extent
networks and content providers have engaged with RPKI. The complexity of this question is
particularly exacerbated by the unusual way websites are often hosted. Generally, they rely on
third party networks to host their servers with many services spreading their content among
multiple networks and CDNs. Enabling RPKI is therefore not just a matter of flipping a switch.
Although this could dissuade websites from utilising RPKI, it does not reduce its importance.


In this paper, we argue that websites should take a holistic approach to securing their services by
extending security provisions beyond traditional end-to-end concerns. Although there are many third
party dependencies that websites should secure, we focus on BGP due to the recency of several attacks in
this realm. We conduct the first quantitative analysis of the deployment of RPKI by web providers.
We make several worrying discoveries, revealing a lack of awareness of the importance of securing BGP among those hosting
websites. Perversely, this ranges from huge international players such as Google to small regional websites.
We then explore the reasons for this, as well as intuitive steps that should be taken. Our findings can be summarized as follows:
\begin{senumerate}
  \item Less popular websites are commonly better secured than 
  websites with many visitors.

  \item CDNs tend to ignore RPKI, whereas ISPs and webhosters have started RPKI
  deployment.

  \item CDN servers that are placed in third party networks benefit from
  RPKI deployment that these networks perform.

  \item CDN deployment policies are the principle cause for a reduced security level at prominent websites.
\end{senumerate}

The remainder of this paper is structured as follows.
\S~\ref{sec:background} details the problem space.
\S~\ref{sec:methoresults} introduces a first proposal of a broadly
reproducible measurement methodology and \S~\ref{sec:results} presents our
first findings. \S~\ref{sec:discussion} sheds light on business reasons why
operators do not deploy RPKI and sketches a research agenda.
\S~\ref{sec:relatedw} briefly surveys related work.
\S~\ref{sec:conclusion} concludes with our outlook.


\section{Background}
\label{sec:background}

\subsection{Primer on RPKI}

The Border Gateway Protocol (BGP)~\cite{RFC-4271} governs the inter-domain routing of the Internet.
BGP exchanges announcements that advertise the ability to reach IP prefixes between Autonomous Systems (ASes).
An AS represents a set of prefixes that it originates and is identified by unique AS Number (ASN).
Unfortunately, a BGP\- speaker may announce \emph{any} IP prefix, allowing a malicious or misconfigured party to disrupt routing.
To undermine access to a website, a malicious BGP speaker could, for example, advertise the website's IP prefix and ``steal'' traffic
destined to it.


To address the above problem, the Resource Public Key Infrastructure (RPKI)~\cite{RFC-6480} has been designed.
It is a PKI framework dedicated to securing the Internet routing infrastructure.
It uses cryptographic certificates to prove the ownership of Internet number
resources (\ie ASNs and IP prefixes).
However, the Internet decouples the ownership of resources and their routing.
Hence, the RPKI introduces Route Origin Authorization (ROA) objects.
A correctly signed ROA authorizes an AS to originate one or more~prefixes.



Using ROA data, an RPKI-enabled router is able to validate the BGP updates it
receives.
There are three states for BGP updates when validated using the RPKI, valid, invalid, and not found.
Rejecting an invalid route announcement helps to suppress incorrectly announced prefix, thus
preventing route hijacking of websites for example. Of course, invalid announcements do not
necessarily indicate malicious behaviour, \eg it could be misconfiguration of ROAs \cite{wms-tdbrh-12}.
However, from a website operator's perspective, this is irrelevant.



\subsection{RPKI and the Web}

To successfully access a webpage, there are usually two key requisites: \one~retrieve a valid domain$\rightarrow$IP address mapping via DNS; and \two~establish valid end-to-end routing from the client to the mapped IP address. DNSSEC~\cite{RFC-2535} can ensure valid name to address mapping. This paper focuses on the routing layer.

To secure the routing layer (via RPKI), a website operator must take certain steps. In the simplest case, a web page is hosted on a single web server situated in a single AS. The prefix owner therefore needs to create a single RPKI entry for the prefix-AS pair, which hosts the domain. However, highly popular webpages are often distributed among several web servers to increase availability and performance. With the advent of Content Delivery Networks (CDNs), these web servers are not only reachable via different IP prefixes but also placed in different ASes. To fully secure the web server infrastructure of the domain, all prefix-AS pairs need to be included into the RPKI. It is worth noting that content provided by CDNs is not necessarily located in the AS of the CDN, in which case the CDN has no control over the authorization of this AS.

\subsection{Attacker Model: Beyond DoS}

Having clarified that the current Web ecosystem complicates the success of countermeasures to correctly access content on the network layer, we now introduce an attacker model which lacks sufficient attention in our community.
We assume an attacker who is able to redirect network traffic destined to the web server by manipulating Internet routing, \ie sending malicious route updates related to the web server infrastructure.
Compared to very common DoS attacks against web servers, this threat might seem less likely.
However, even difficult attacks find their way into the real world, sooner or later.
A simple example we refer to the Great Cannon attack beginning this year~\cite{mwdef-cgc-15}, where an ISP injected on-path malicious JavaScript code into live network traffic to disturb connectivity to GitHub.
Note that TLS does not necessarily protect against such an attack when prefix hijacking is in place \cite{g-bhwbh-15}.

There are political as well as commercial reasons that motivates an attacker to attack a web server.
Leveraging prefix hijacking has several attractive advantages.
First, the attacker can intercept the network traffic, which enables him to drop, monitor, or modify packets.
Second, the attack does not necessarily need to affect all clients---when malicious route propagation is restricted locally, the attacker can harm specific subset of clients.
Third, the attack can be performed without notifying the website under attack. In the following, we explore the severity of this threat and the level of countermeasures currently deployed via RPKI.




\section{Preliminary Methodology}
\label{sec:methoresults}

Our measurement study proceeds in four steps: (1) selection of websites; (2) mapping domain names to IP addresses; (3) mapping these IP addresses to IP prefixes routed in the Internet and their origin ASes; and (4) validating if these ASes support RPKI. The purpose is to explore how widely deployed RPKI is among web providers. We now describe our methodology, which is meant to be simple and widely reproducible. All data will be made available.

\paragraph*{(1) Selecting Domain Names} 

To explore RPKI deployment among websites, it is necessary to first select a sample set. To achieve this, we extract the top 1 million websites from the Alexa list~\cite{alexa-ranking}. This is the default approach taken by most web measurements (\eg \cite{amsu-wcc-11,rkw-ddatt-12,lgs-bspdj-13,ctbo-qmclm-14}), and allows us to explore how RPKI deployment varies across popularity~ranks.


\paragraph*{(2) Mapping Domains to IP Addresses}

To map domain names taken from the Alexa into IP addresses, we utilise several public resolvers (to improve reproducibility).
Using Google DNS, we collect all \texttt{A}, \texttt{AAAA}, and \texttt{CNAME} records for all the Alexa domain names, including the names appended
with the prefix \url{www}. We refer to the former as a ``w/o www domain''. The measurements were performed from Berlin repeatedly over several weeks in 2014 and 2015.
We also used Open DNS and \url{us01} of the DNS Looking Glass~\cite{dns-lg} to verify that the results returned from Google DNS were not miscellaneous.
We obtain similar results from all three datasets.
We exclude all invalid DNS answers, \ie all
special-purpose IPv4 and IPv6 addresses reserved by the IANA.

Note that, although the distributed nature of DNS may lead to  answers that vary between locations, we delay exploration of CDN redirection strategies to future work.
This is because we are primarily interested in checking whole ASes, rather than where individual clients are redirected to.
In Section~\ref{sec:results}, we show that our main results remain independent of the DNS server selection because CDNs are reluctant to create ROAs at all.
Analyzing the effects of many vantage points will be part of our future work.

\begin{figure}
  \includegraphics[width=\columnwidth]{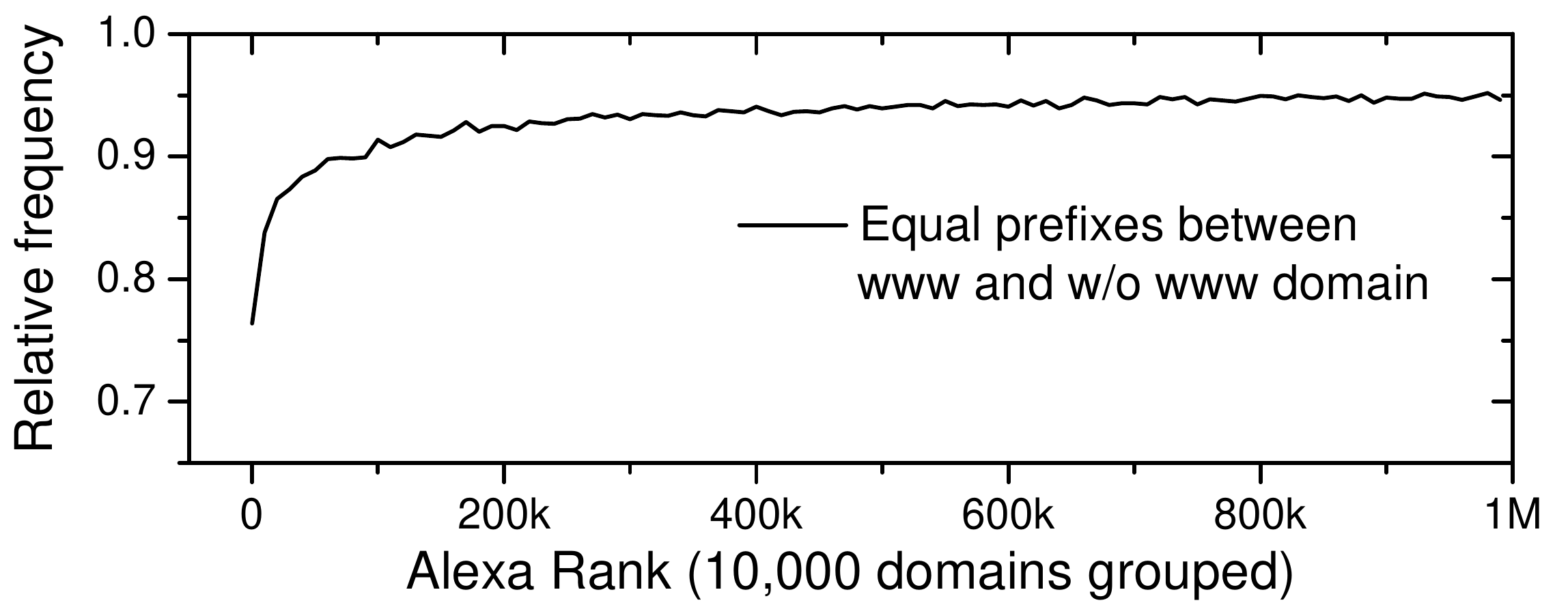}
  \caption{Comparison of IP deployment for www and w/o www domain names.}
  \label{fig:prefixmatch-www}
\end{figure}

To briefly analyze the overlap between www and w/o www domains, we quantify
the amount of equal prefixes per domain. Figure~\ref{fig:prefixmatch-www}
shows that for the first 100k domains more than 76\% of the IP prefixes are
equal for both names. For the remaining domains, more than 94\% of the names
refer to the same prefix. As a side observation, in future work it should be explored how this fact can help accelerate continuous DNS measurements.

\paragraph*{(3) Mapping IP Addresses to Prefixes and ASNs}
To map addresses to ASNs, we take dumps of the active tables of the RIPE RIS route servers. For each IP address of a domain
name, we extract all covering prefixes and derive the origin AS from the AS
path (i.e., the right most ASN in the AS path). Note that entries with an
AS\_SET are excluded from our study as this leads to an ambiguity of the attribute, which is why the 
function is deprecated with the deployment of RPKI~\cite{RFC-6472}.

\paragraph*{(4) RPKI Validation}
For the validation of the BGP data, we follow the necessary steps to
perform origin validation at BGP routers. ROA data of all trust anchors (APNIC, AfriNIC, ARIN, LACNIC, and
RIPE) are collected and validated. Only cryptographically correct ROAs are
further used to check the IP prefixes obtained 
from the BGP table dumps.

The outcome of this process is a comprehensive list of all Alexa websites that \one can be resolved from our DNS vantage point and \two mapped to an IP prefix AS pair.
This list is \three annotated with RPKI origin validation outcome.


\section{Results}
\label{sec:results}

We have used the above methodology to collect data on RPKI deployment for all 1 million Alexa domains.
After excluding 0.07\% incorrect DNS answers, we gathered 1,167,086 IP addresses
for the www domains and 1,154,170 IP addresses for the w/o www domains.
These addresses map to 1,369,030 and 1,334,957 different prefix-AS
pairs respectively. 0.01\% of the IP addresses are not reachable from our
BGP vantage~points.

In the following subsections, we present the core RPKI
validation outcomes, and explore reasons for the observed deployment
state. For better visibility, we do not present results per domain but
apply a binning of 10k domains in all graphs, after experimenting with different bin sizes. As a domain name may refer to multiple IP
addresses, which may belong to different IP prefixes and ASes, several RPKI
states may exist per domain. To represent heterogeneous RPKI deployment, we assign corresponding probabilities to domain names (\eg 3/5 or 60\% RPKI coverage of \url{foo.bar}).


\subsection{Less Popular Content is More Secured}

\begin{figure}
  \includegraphics[width=0.49\textwidth]{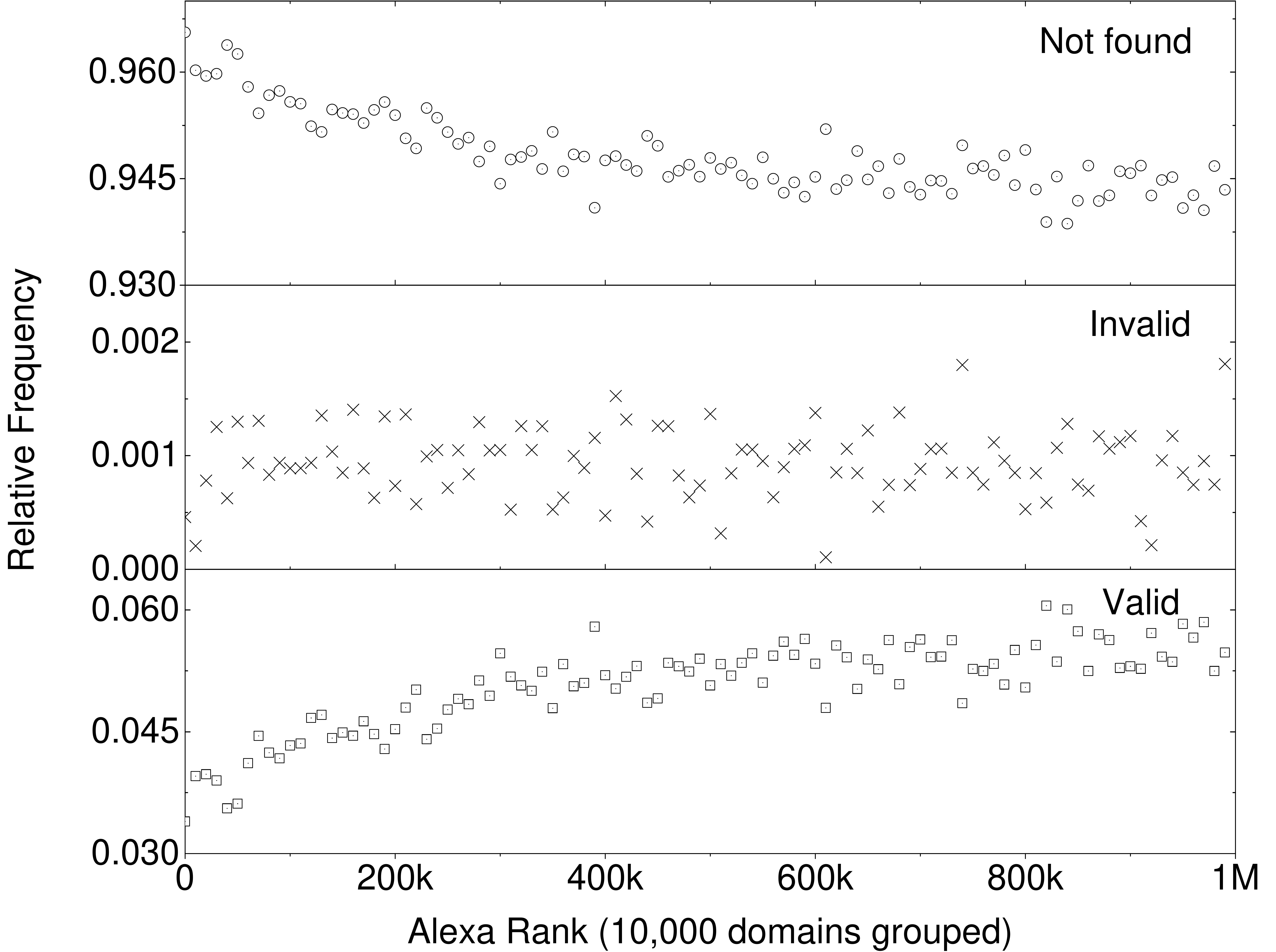}
  \caption{RPKI validation outcome for the 1 million Alexa domains (\emph{valid} =
  the origin AS is allowed to announce the prefix, \emph{invalid} =  the
  origin AS is not allowed to announce the prefix, and \emph{not found} =
  the announced prefix is not covered by the RPKI).}
  \label{fig:rpki-validation}
\end{figure}

We first explore how many websites are secured via RPKI, shown in Figure~\ref{fig:rpki-validation}.
On average, only 6\% of the web server prefixes are covered by RPKI (either
correctly or incorrectly announced in the BGP). This is somewhat worrying considering the international prominence
of the websites under study.
However, it also effectively illustrates that web operation and network operation are decoupled.
Roughly 0.09\% of the
prefixes are invalid according to the RPKI prefix origin validation, spread evenly across all Alexa ranks. This
observation is in qualitative agreement with the general RPKI deployment. Note
that the current invalid BGP announcements do not necessarily indicate
hijacking, but rather potential misconfiguration \cite{wms-tdbrh-12}. The amount of
invalids is evenly distributed among all web domains.

Perhaps more interesting is the distribution of RPKI support across the different popularity rankings.
By inspecting Figure~\ref{fig:rpki-validation}, distinct trends
can be seen. 
Perversely, domains with a high rank (\ie
popular sites in the top 100k) are less likely to be secured than less popular sites.
Among the first 100k domains (\eg \url{google.com}, \url{blogspot.com}), only $\approx$4.0\% of web server prefixes are
secured via RPKI.
In contrast, for the last 100k domains, $\approx$5.5\% are secured. The absolute numbers are small, but a clear trend is visible and may
reflect the deployment strategy of different stakeholders.

Table~\ref{tbl:top10rpkisites} shows the first top 10 domains that have RPKI support, \ie either all or at least one prefix which is associated with the domain name is part of the RPKI.
It is clearly visible that \one almost all of the very popular sites are unsecured; and \two sometimes there is differing RPKI support for a website's www and w/o www domains; and \three most of the content is only partially secured.
We now analyze the deployment by CDNs in more detail.

\begin{table}
\center
\small
\begin{tabular}{rlcc}
\toprule
& & \multicolumn{2}{c}{\textbf{RPKI Coverage}}
\\
\cmidrule{3-4}
\multicolumn{2}{l}{\textbf{Alexa Rank \& Domain Name}} & \textbf{www} & \textbf{w/o www}
\\
2 & facebook.com & \cmark\xspace(3/3) & \cmark\xspace (2/2)
\\
70 & cdncache1-a.akamaihd.net & n/a & \pmark\xspace (1/3)
\\
73 & huffingtonpost.com & \pmark\xspace (1/3) & \xmark\xspace (0/3)
\\
92 & cnet.com & \pmark\xspace (1/3) & \xmark\xspace (0/2)
\\
95 & dailymail.co.uk & \pmark\xspace (1/3) & \xmark\xspace (0/1)
\\
117 & indiatimes.com & \pmark\xspace (1/3) & \xmark\xspace (0/1)
\\
120 & kickass.to & \ \pmark\xspace (1/10) & \ \pmark\xspace (1/10)
\\
130 & booking.com & \cmark\xspace (4/4) & \cmark\xspace (2/2)
\\
\bottomrule

\end{tabular}
\caption{Top~10 Alexa domains that have partial (\pmark) or
full (\cmark) RPKI coverage, including number of prefixes.}
\label{tbl:top10rpkisites}
\end{table}


\subsection{CDN Content Benefits from 3\textsuperscript{rd} Party ISPs}

CDNs are a critical part of modern web delivery. Many websites rely on them to deliver their content.
Next, we focus on the RPKI deployment of CDNs. To study this, we search the RPKI repository for attestation
objects that belong to the ASes of well-known CDNs. Specifically, we inspect the infrastructures of Akamai, Amazon,
Cdnetworks, Chinacache, Chinanet, Cloudflare, Cotendo, Edgecast, Highwinds,
Instart, Internap, Limelight, Mirrorimage, Netdna, Simplecdn, and Yottaa.
It is worth noting that the results of this approach do \emph{not} depend
on DNS measurements and thus do not include a bias that might
result from any DNS measurement point.





To derive the AS numbers of these CDNs, we apply keyword spotting on common AS assignment lists.
This leads to a lower bound for the current state of deployment.
We discover 199 ASes operated by these CDNs.
From these, we find only four entries in the RPKI.
These four prefixes are owned by Internap and are tied to three origin ASes. One might mistakenly think that Internap has therefore engaged widely with RPKI. However, Internap operates at least 41 ASes, the bulk of which are not secured via RPKI. No other CDN has made any deployment. Thus, these CDNs do not actively participate in the creation of RPKI attestation objects. This is in contrast to web hosters or common ISPs that, as shown previously, have far higher levels of penetration ($>5\%$).


Another interesting trend has been for CDNs to place caches in third party networks (\eg eyeball ISPs).
This allows the CDN to ``inherit'' RPKI support from the third party network.
Based on our previous observation, \ie these CDNs do not create ROAs, we know that every RPKI-enabled CDN-content is served by a third party network.


\subsection{Are the CDNs to blame?}

To quantify the impact that CDNs have on global website support for RPKI, we conduct a basic classification of CDN domains.
Generally, CDNs use CNAME chains (Canonical Names) to redirect DNS requests to
their caches. We exploit this fact to identify how many of our Alexa domains use CDNs.
We say a domain is served by a CDN, if the IP address
of its domain name is indirectly accessed via two or more CNAMEs (\eg
\url{www.huffingtonpost.com} $\rightarrow$
\url{www.huffingtonpost.com.\\edgesuite.net} $\rightarrow$
\url{a495.g.akamai.net} $\rightarrow$ \url{212.201.100.136}).
We confirm this rough heuristic by comparing its results with an independent
classification provided by HTTP\-Archive. HTTP\-Archive classifies the first
300k Alexa domains based on DNS pattern matching of CNAMEs, which is distinct from our test of DNS indirections. Furthermore, the HTTPArchive monitoring agent is located in Redwood City, CA, USA, and thus a geographically separated vantage point.

\begin{figure}
  \includegraphics[width=\columnwidth]{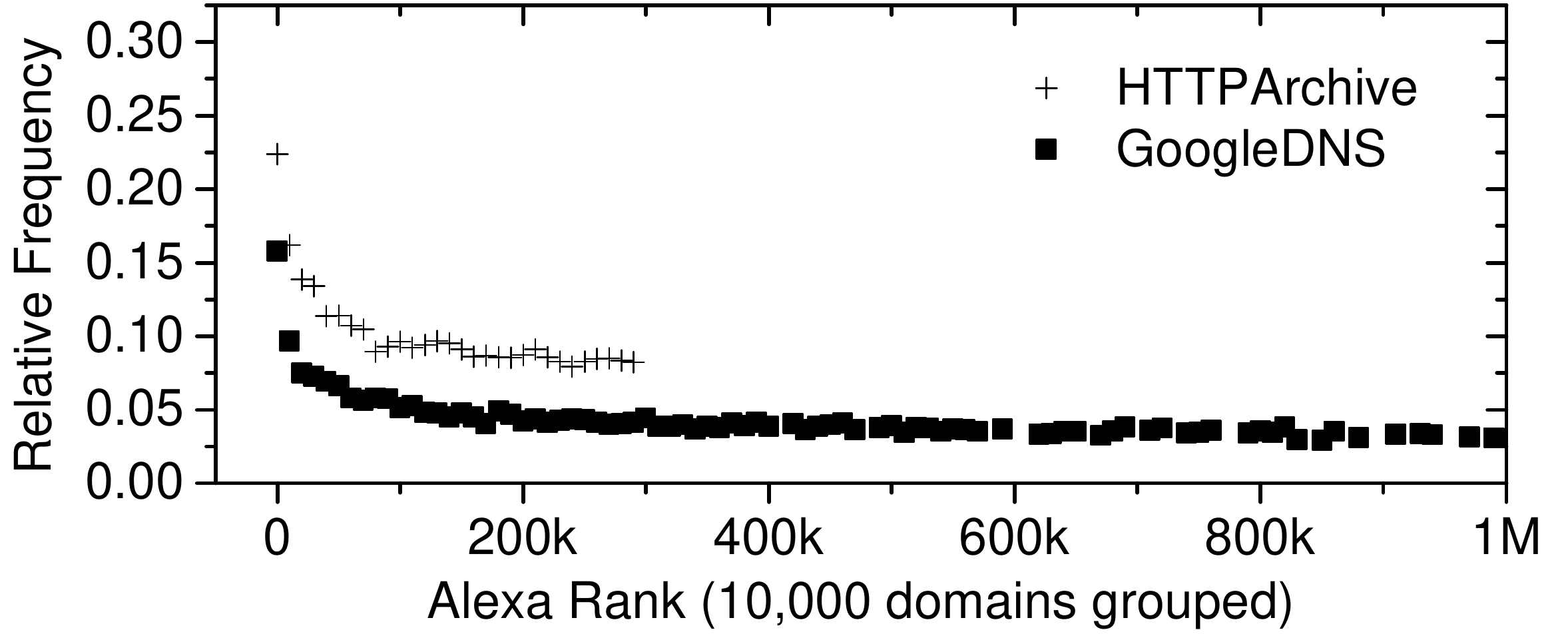}
  \caption{Popularity of CDNs---comparison of CDN detection heuristics for 1M Alexa domains.}
  \label{fig:cnameprob-httparchive-google-www}
\end{figure}

Figure~\ref{fig:cnameprob-httparchive-google-www} compares the distributions of 
CDN-hosted web domains as determined by our classification approach and  
HTTPArchive (in 10k bins). The two almost identically shaped curves clearly
indicate that popular websites are more likely to be served by CDNs. 
Quantitatively, our approach indicates fewer CDNs than HTTPArchive. This is not surprising, since there are CDN deployments without CNAME chains. However, a conservative (un\-der)-estimate of CDN domains sharpens our view on the RPKI-protection of CDN domains: (Over-)en\-larg\-ing the set of CDN domains will mix deployment cases and diffuse the overall picture.

\begin{figure}
  \includegraphics[width=\columnwidth]{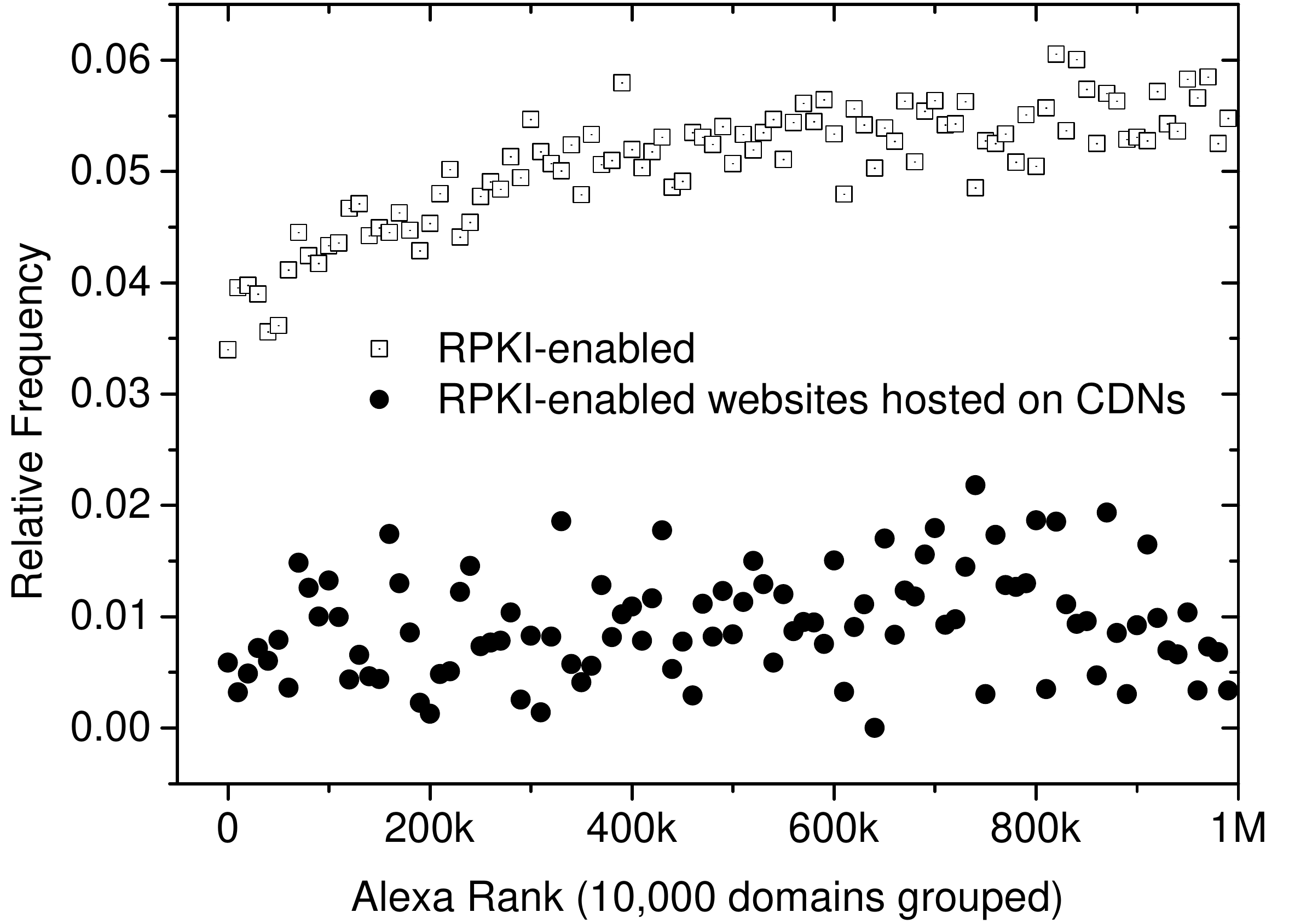}
  \caption{RPKI deployment statistics on CDNs and for the unconditioned Web.}
  \label{fig:rpki-cdn-prob}
\end{figure}

Integrating the above results, we now explore why highly ranked websites do not support RPKI. Specifically, we focus on the relation between RPKI-enabled and CDN-served websites.
Figure~\ref{fig:rpki-cdn-prob} depicts the distribution of RPKI-enabled websites across the Alexa rankings. The figure separates websites into those that utilise a CDN and those that do not.
In contrast to the Alexa domains at large, RPKI deployment is fairly independent of the rank for CDNs. Results fluctuate around an average of $\approx 0.9\%$. This is almost an order of magnitude lower than the overall RPKI deployment rate, which is plotted for comparison.

Combining the lines of argument, we have shown that \one~CDN deployment is strongly enhanced for popular domains, but \two~RPKI deployment for CDNs is low---independent of content popularity. As a result, a high density of CDN sites reduces the RPKI-enabled portion of domains.
This holds for those ranks of the Alexa list where CDNs are more common: at the top ranks.
In summary, the observable degradation of routing security for popular websites is caused by the resistance of CDN operators to adopt RPKI in their ASes.


\section{Discussion and Research Agenda}
\label{sec:discussion}

After \one identifying a new attacker model, which has substantial potential to threaten the Web, and \two analysing the current deployment of the state-of-the-art countermeasure, we are left with a surprisingly basic but still unanswered question: How can a content owner easily verify that his content is reliably and securely delivered in the current Web ecosystem?

\subsection{Securing Web Architectures}

First and foremost, we call for a more elegant and approachable web architecture. We currently see a complex ecosystem built by a broad set of (sometimes competing) stakeholders. These include many network operators, content providers, and CDNs, largely underpinned by an abused DNS infrastructure. At each stage in the content delivery chain, different security protocols are being employed, making a comprehensive understanding virtually impossible for most web startups. For instance, a key finding of our work is the poor awareness that web providers have regarding routing security. Despite its importance, only a small fraction of websites have any deployment of RPKI.

The above observations are unlikely to change in the near future as the deployment of such mechanisms require notable effort. As of yet, we have not witnessed events critical enough to shake most web providers into action. However, we should not be sitting and waiting for a major event to disrupt the Internet content delivery ecosystem, as we have already many examples hinting at its potential impact~\cite{sb-arc-10,bitcoin_hijack,youtube-outage}. A key part of our future research agenda is expanding our work from RPKI to include other security mechanisms that web providers should be aware of (\eg DNSSEC). Ultimately, we argue that new systems should be devised that increase transparency and thus make the comprehensive understanding and deployment of e2e security much more straightforward. 


\subsection{Incentivizing RPKI Deployment}

Returning to our key contribution, we also argue that the need for RPKI must be more critically communicated to web providers---particularly large CDNs that serve disproportionate amounts of web traffic. This is largely a process of minimising barriers (\eg cost), alongside offering incentives. During our study, we have spoken to many network operators to gauge their opinion of RPKI.
We have found several reasons why operators have not deployed it (\eg \cite{g-arn-15}). Often this relates to a lack of perceived need, combined with insufficient manpower and expertise in the area. This is common across many technologies. More interestingly, we also have discovered RPKI-specific factors that dissuade adoption.

RPKI is a proactive security solution that implements a positive attestation model.
As such, RPKI exposes information that organisations may be wary of revealing.
Most worrying to some is the potential of revealing their business
relations.
In RPKI, prefix owners must proactively create ROAs before any attack occurs.
As soon as at least one ROA for an IP prefix exists, \emph{all} valid origin ASes for this IP prefix need to be assigned in the RPKI \emph{before} route updates are processed (otherwise a BGP update including the prefix and missing ASes becomes invalid).
Although a prefix owner can assign any AS without asking for approval, it is very likely that the ROA information indicates a business relation between prefix owner and authorized origin AS.
This might be a serious concern, from our discussions with several operators.

To illustrate a business policy conflict, imagine that two large CDNs serve secretly as backups for each other.
As changes within the DNS are slow, they could quickly redirect traffic using BGP.
Similarly, smaller CDNs that rely on third party networks (\eg using Verisign for external DoS mitigation) may see such information as damaging to their reputation. Despite this, in both cases, RPKI would publicly reveal these setups.


It is worth noting that RPKI data differs from public routing data such as BGP collectors or looking glasses.
Those sources also provide insights into peering relations but only after the event has occurred.
Furthermore, the data analysis follows an exploratory approach because not all vantage points report the same.
In contrast to this, the RPKI represents a catalog which does not only allow for easy browsing but also documents information in advance.
In case of a very unlikely or never occurring event (e.g., a backup incident), the RPKI exposes more information.
We therefore argue for changes to RPKI that address these business concerns. Whereas privacy-aware protocols are rife in other fields (\eg Tor), they have always been seen as less important in the routing layer. The above observations undermine this assumption. Arguably, not recognising this issue could be extremely damaging to RPKI deployment.

\subsection{Web Measurement Methodology}

A tangential observation from our work is the daunting complexity of launching web measurement campaigns. Compiling a generalised and comprehensive methodology for dealing with this still eludes the web community. Questions that remain open include: \one~which websites should be queried; \two~where should they be queried from; \three~how these should be distributed over time. For instance, selecting Alexa websites is a typical approach, however, results differ for www and w/o www domains; complexity is also greatly increased when considered the tendency to shard content across multiple subdomains in a website. Some might say that security incidents in the past~\cite{youtube-outage,indonesia-hijack,sb-arc-10} have showed that routing failures usually affect complete web pages instead of just subdomains. However, a commercially motivated attacker may explicitly target subdomains, \eg those hosting adverts. Exhaustive crawls may be suitable for small sets of websites, but scalability quickly becomes a challenge particularly when duplicating crawls across regions. Answering these questions at scale is not trivial, and should be explored further. We plan to use such insights to expand our methodology in a generalised way that can help others to perform web measurements too. Without better methodologies in this field, web measurements may become increasingly infeasible and unreproducible as web complexity increases.


%
%

\section{Related Work}
\label{sec:relatedw}

The deployment of RPKI started in 2011. Several looking glasses and tools
exist
\cite{lacnic-rpki-lg,surfnet-rpkidashboard,nist-rpkimonitoring,knmpn-asene-13,whss-roslr-13,rws-rmmir-15b}
to inspect the current state of deployment or to do experiments, but up
until now only few publications studied the current state of deployment in
detail. \cite{wms-tdbrh-12,ipb-mbror-15} analyze the RPKI validation outcome of entire
BGP tables trying to better understand invalid BGP announcements.
\cite{chbrg-rmra-13} discusses the risk when RPKI authorities misbehave,
and \cite{gshr-ssirp-10} explores the general limitations of current secure
inter-domain routing protocols.  Nevertheless, large ISPs such as Deutsche
Telekom and ATT added their IP prefixes to the RPKI, which motivates the
relevance of this new protocol framework. The motivation to adapt new
Internet protocols is analyzed in \cite{rsm-teia-05}, with a special focus on
secure inter-domain routing in \cite{cdpz-masbp-06,gsg-lmdds-11}. Our work
complements these insights by clarifying that CDNs can benefit from early RPKI-adoption in third party networks where their servers are hosted.

Several measurement studies discovered the content distribution space
(\eg \cite{skg-wsts-07,hwlr-melc-08,sckb-dbain-09,amsu-wcc-11}). We
emphasize that the aim of this paper is not to reveal the hosting
infrastructure completely but to present a trend analyse of RPKI adoption
in the wild and its interplay with the web ecosystem by applying a very
basic methodology.

\section{Conclusion and Outlook}
\label{sec:conclusion}

In this paper, we analyzed the RPKI-protection of websites. We resolved 1M
domain names from the Alexa ranking, mapped the IP addresses to IP prefixes
and origin ASes visible in the global BGP routing table, and validate each
prefix-AS pair against the currently deployed RPKI data.

We found that RPKI security deployment is significantly degraded for the more popular websites, which led us to apply a new, initial methodology for discovering its reasons. Our findings revealed that CDN hosters are the likely cause for this operational bias. Their enhanced provenience  at prominent web domains on the one hand, and their obvious reluctance towards RPKI deployment on the other hand strongly indicate that prominent websites would be better protected against routing attacks without CDNs.

In future work, we will aim to explore in more detail why CDNs implement this operational
behavior. Furthermore, we will compare RPKI deployment with the adoption of
other core protocols such as DNSSEC.

\section*{Acknowledgements}
We would like to thank Eric Osterweil and Andy Newton for very fruitful discussions.
This work was partially supported by the BMBF within the project Peeroskop.

\balance
\small
\bibliographystyle{abbrv}
\bibliography{rfcs,ids,internet,security,own,tmp}

\end{document}